\documentclass[aps,prb,twocolumn,amsfonts,amssymb,amsmath,floatfix,showpacs,nobalancelastpage,raggedbottom]{revtex4-1}
\usepackage{tikz}
\usepackage[strict]{revquantum}
\usepackage[T1]{fontenc}
\usepackage{times}
\usepackage{epstopdf}
\usepackage{graphicx}
\usepackage{bm,euscript}
\usepackage{color}
\bibpunct{[}{]}{,}{n}{}{}
\begin{document}

\title{Many-Body Localization in  Spin Chain Systems with  Quasiperiodic Fields}

\author{Mac Lee}
\author{Thomas R. Look}
\author{D. N. Sheng}
\author{S. P. Lim}

\affiliation{Department of Physics and Astronomy, California State University, Northridge, California 91330, USA}

\begin{abstract}
We study the many-body localization of spin chain systems with quasiperiodic fields. We identify the lower bound for the critical disorder necessary to drive the transition between the thermal and many-body localized phase to be $W_{cl}\sim 1.85$, based on finite-size scaling of entanglement entropy and fluctuations of the bipartite magnetization.  We also examine the time evolution of the entanglement entropy of an initial product state where we find power-law and logarithmic growth for the thermal and many-body localized phases, respectively.  For larger disorder strength, both imbalance and spin glass order are preserved at long times, while spin glass order shows dependence on system size.  Quasiperiodic fields have been applied in different experimental systems and our study finds that such fields are very efficient at driving the MBL phase transition.
\end{abstract}

\pacs{73.40.Hm, 71.30.+h, 73.20.Jc}

\maketitle

\section{Introduction}

The interplay of random disorder with many-body interactions has attracted a lot of recent research activity\cite{basko2006, oganesyan2007, pal2010, znidaric2008, huse2013, nandkishore2015, altman2015, huse2014, nandkishore2014, pekker_hilbert2014}.  The many-body localized (MBL) quantum phase\cite{nandkishore2015, altman2015, huse2014, nandkishore2014, pekker_hilbert2014} of matter is distinctly different from both Anderson localized phase  in non-interacting systems and the ergodic (thermal) phase for interacting systems with weaker disorder.  The ergodic phase follows the  eigenstate thermalization hypothesis (ETH) which describes how an isolated self-interacting quantum system can thermalize under its own internal dynamics in agreement with quantum statistical mechanics\cite{deutsch1991, srednicki1994, rigol2008}.  A system in the MBL phase, on the other hand, fails to thermalize even for its highly excited eigenstates on any time scale, resulting in new statistics for such systems\cite{basko2006, oganesyan2007, pal2010, znidaric2008, huse2013, nandkishore2015, altman2015, huse2014, nandkishore2014, pekker_hilbert2014}.  Many remarkable properties of the MBL phase have been established\cite{nandkishore2015, altman2015, huse2013, nandkishore2014, oganesyan2007, pal2010, znidaric2008, rigol2008, serbyn2014, kwasigroch2014, yao2014, vasseur2015, huse2014, serbyn2013, ros2015, chandran2014, grover2014, agarwal2015, knap2015, luitz2015, devakul2015, torres2015, canovi2011, cuevas2012, bauer2013, kjall2014, luca2013, iyer2013, pekker_hilbert2014, johri2014, bardarson2012, andraschko2014, laumann2014, hickey2014, nanduri2014, barlev2014, imbrie2014, groverf2014, ponte2015, huang2015, you2015, serbyn2015, singh2015, barlev2015, deng2015, chen2015} based on extensive theoretical studies.  The existence of both the ergodic and MBL phases dictates  a novel dynamic quantum phase transition between them\cite{basko2006, pal2010, oganesyan2007, kjall2014, vosk_theory2014, potter2015trans, serbyn2015, agarwal2015, knap2015,baygan2015,zhang2016, zhang2016a, yu2016, vedika2016, dumitrescu2017}.  Random disorder introduces rare Griffiths regions\cite{vosk_theory2014, potter2015trans, knap2015,luitz2015,khemani2015,yu2015,lim2016,kennes2015} which may have singular contributions in driving such a phase transition, but there is still a limited quantitative understanding of their effects.  Quasiperiodic fields\cite{aubry1980} have a period incommensurate with the lattice constant, thus they break translational invariance and introduce disorder in a more controlled way when compared to random fields.  

In a recent work, it was shown that interacting quasiperiodic models can have an MBL phase\cite{iyer2013}, and signatures of this phase have been experimentally observed in recent cold-atom experiments\cite{schreiber2015, bordia2015, bordia2016, luschen2016a, luschen2016b}.  However,  most numerical studies of the MBL transition have focused on models of spin chains with random fields\cite{pal2010, kjall2014, luitz2015, yu2016, vedika2016}.  Very recently, the dynamic quantum phase transition has been analyzed\cite{vedika2017,newpaper, nag2017, barlev2017} for systems with quasiperiodic potentials.  By analyzing the intra-sample and inter-sample fluctuations with a close comparison between quasiperiodic and random  fields, Khemani, et al.\cite{vedika2017}, have demonstrated the possibility of two universality classes for the quantum phase transition\cite{vedika2017, vedika2016}. Other studies explore the interplay of MBL in quasiperiodic potentials and the single-particle mobility edge\cite{modak2015,deng2016,nag2017}.  Time evolution of many-body systems have been studied for spin-chains with  randomly distributed fields\cite{kjall2014,luitz2016time} and quasiperiodic fields\cite{newpaper}, which can be used to address the dynamics of the thermal to MBL phase transition\cite{nandkishore2015,vosk_theory2014,potter2015}.  After a global quantum quench, the power-law growth of bipartite entanglement entropy is observed for thermal states while logarithmic growth is found  for MBL states where  local memories of an initial product state persist for all time\cite{luitz2016time,barlev2017}.

In this paper we report on eigenstate and time-dependent studies of spin chains with quasiperiodic fields.  Through exact diagonalization (ED) and Lanczos  Krylov space time evolution calculations, we find a dynamic quantum phase transition from the ergodic phase to the MBL phase that is similar to spin chains with random disordered fields.  However, systems with quasiperiodic fields appear to be more efficient at localizing quantum states which is demonstrated by a smaller critical disorder, $W_{cl} \sim 1.85$ (as a lower bound) compared to similar estimate for systems with random fields\cite{luitz2015} (where the critical disorder field strength is around 3.5\cite{luitz2015}) in agreement with the work of Khemani, et al.\cite{vedika2017}  We also evolve a randomly selected initial product state and study how entanglement entropy and other observables behave as a function of time.  Similar to random field systems, we find that bipartite entanglement entropy experiences power-law growth in the thermal phase and logarithmic growth in the MBL phase.  Interestingly, we also observe quasiperiodic oscillations of spin imbalance on short timescales.  Preservation of imbalance and spin glass order at long times is characteristic of the MBL phase, commensurate with stronger disorder.  Our results suggest a critical quasiperiodic field strength of $W_c\sim 2.5$, and provide a quantitative understanding of the MBL phase for spin systems with quasiperiodic fields.

\section{Theoretical model and  ergodic to many-body localized phase transition}

We study the Heisenberg spin-1/2 chain with a quasi-periodic field
\begin{equation}
H = J\sum_{i=1}^{L-1} \mathbf{S}_i \cdot \mathbf{S}_{i+1} + W\sum_{i}^{L} \cos(2\pi c i+\phi) S_i^z \text{.}
\end{equation}
where $\mathbf{S}_i$ is the spin operator for site $i$, $J$ is the nearest neighbor coupling constant which we set to $J=1$, $W$ is the strength of the quasi-periodic field, $c$ is an irrational wave number chosen to be $c=\sqrt{2}$, and $\phi$ is a random phase used to create different quasiperiodic field  configurations. $L$ is the number of sites (system length).  This model is similar to the one studied recently in \cite{vedika2017}, which  included second nearest neighboring transverse spin couplings.  In this paper, however, we focus on the time-evolution of initial product states for systems with quasiperiodic fields.  We use open-boundary condition which allows for a larger window to observe the time evolution of physical quantities\cite{luitz2016time} before they saturate due to finite-size effects.

\begin{figure}[t]
	\includegraphics[angle=-90,width=0.49\linewidth]{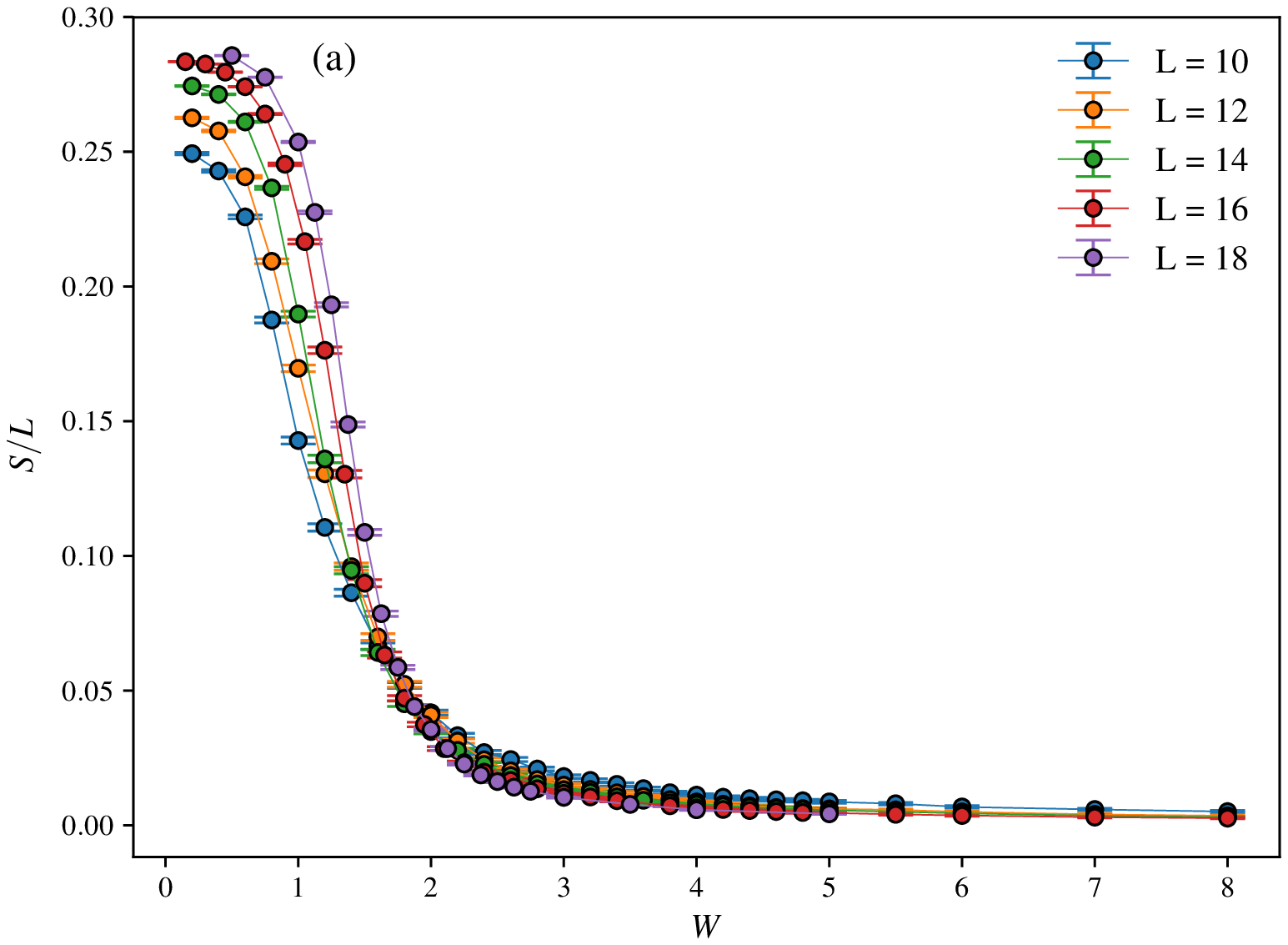}
	\includegraphics[angle=-90,width=0.49\linewidth]{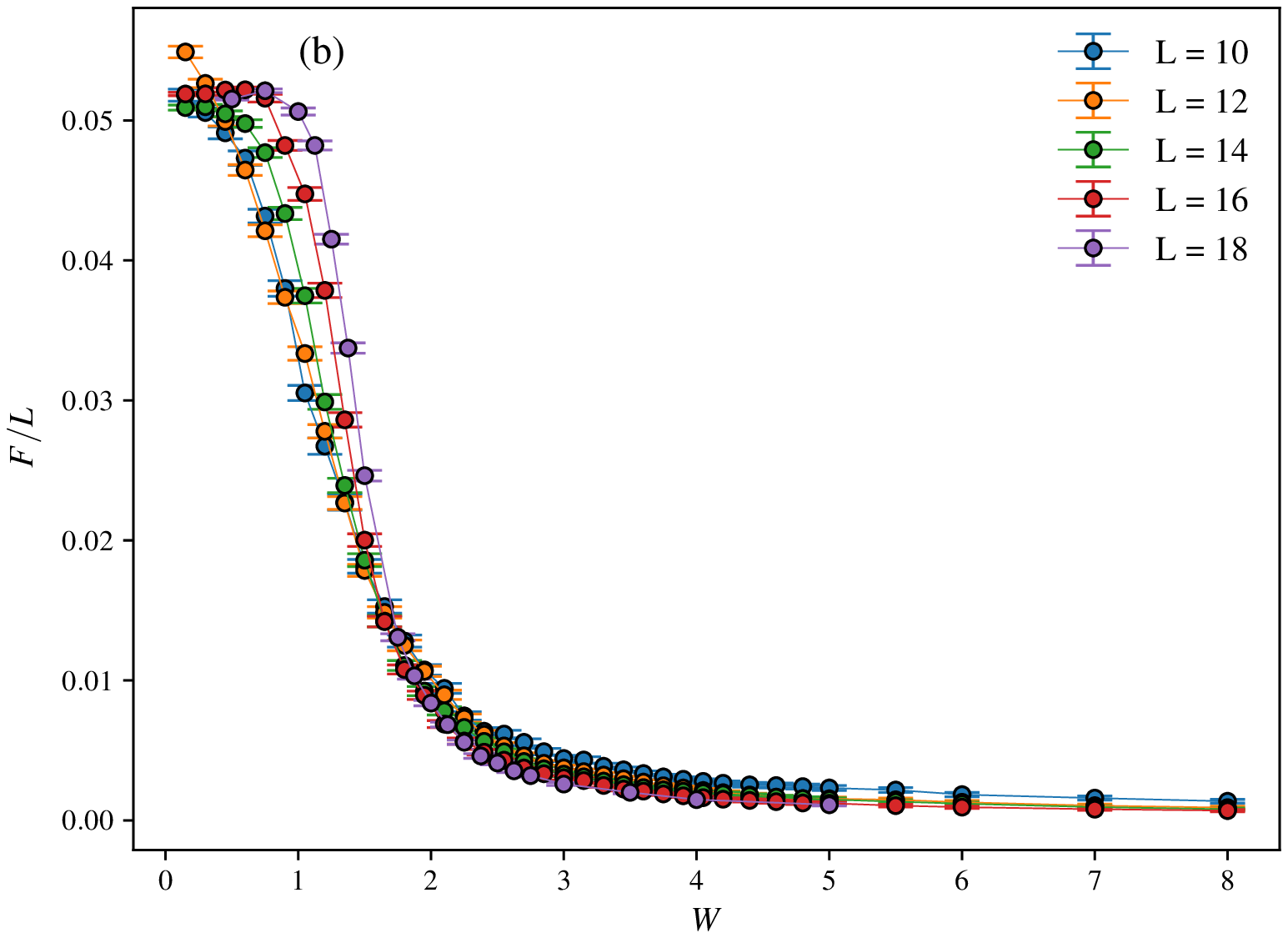}\\
	\caption{
		(Color online) 
		(a) The ratio of entanglement entropy over the number of system sites $S/L$ for $L=10-18$ at the energy density $\varepsilon=0.5$ as a function of the  strength of the quasiperiodic fields $W$.  (b) 
		The fluctuations of the half system magnetization over $L$ for $L=10-18$.  Both graphs display crossing around $W_{cl}\sim 1.85$, suggesting a quantum phase transition at that point.  For larger system sizes, the crossing point drifts towards larger $W$.
	}
	\label{fig1}
\end{figure}

\begin{figure}
	\includegraphics[width=0.49\linewidth]{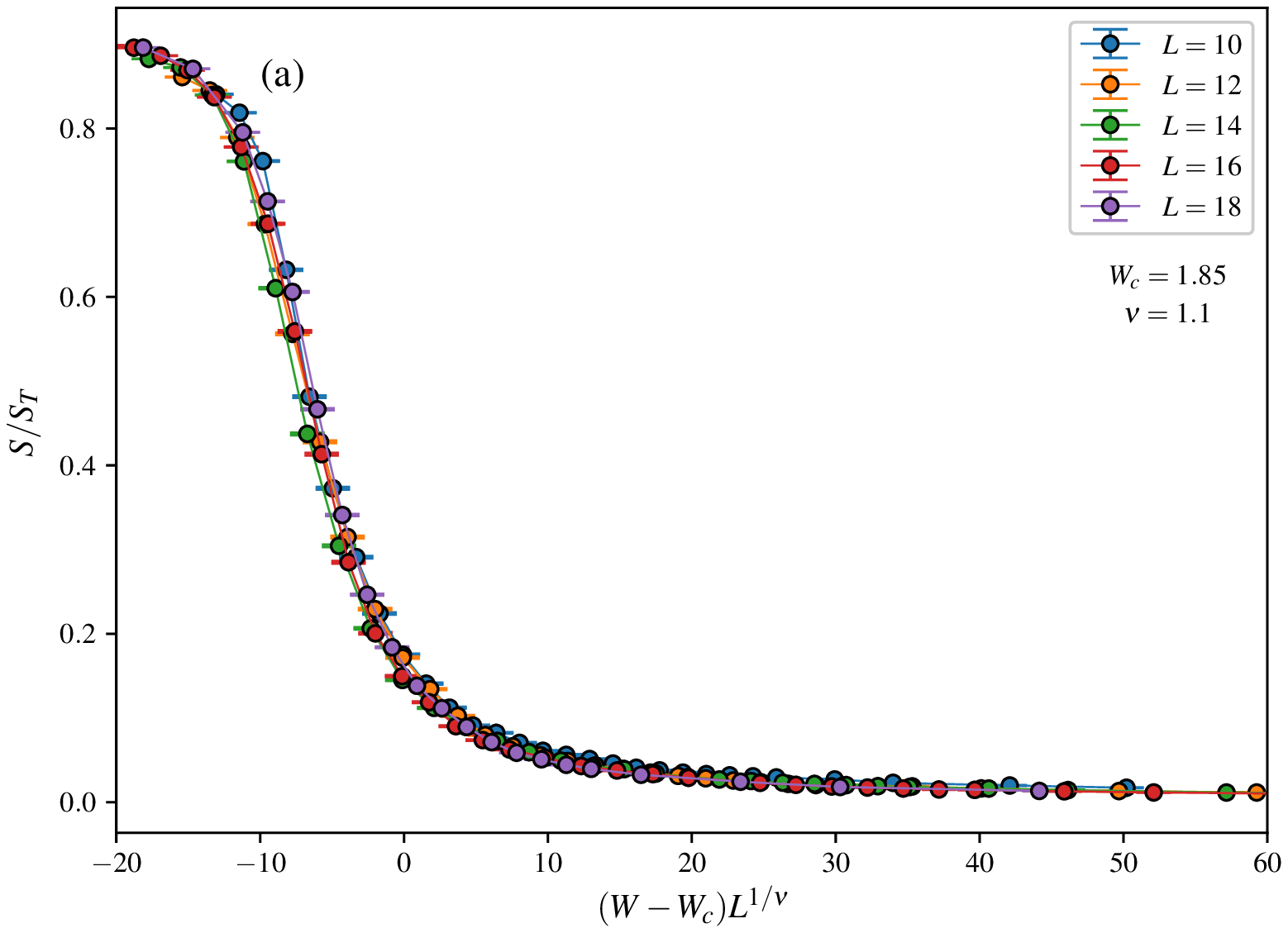}
	\includegraphics[width=0.49\linewidth]{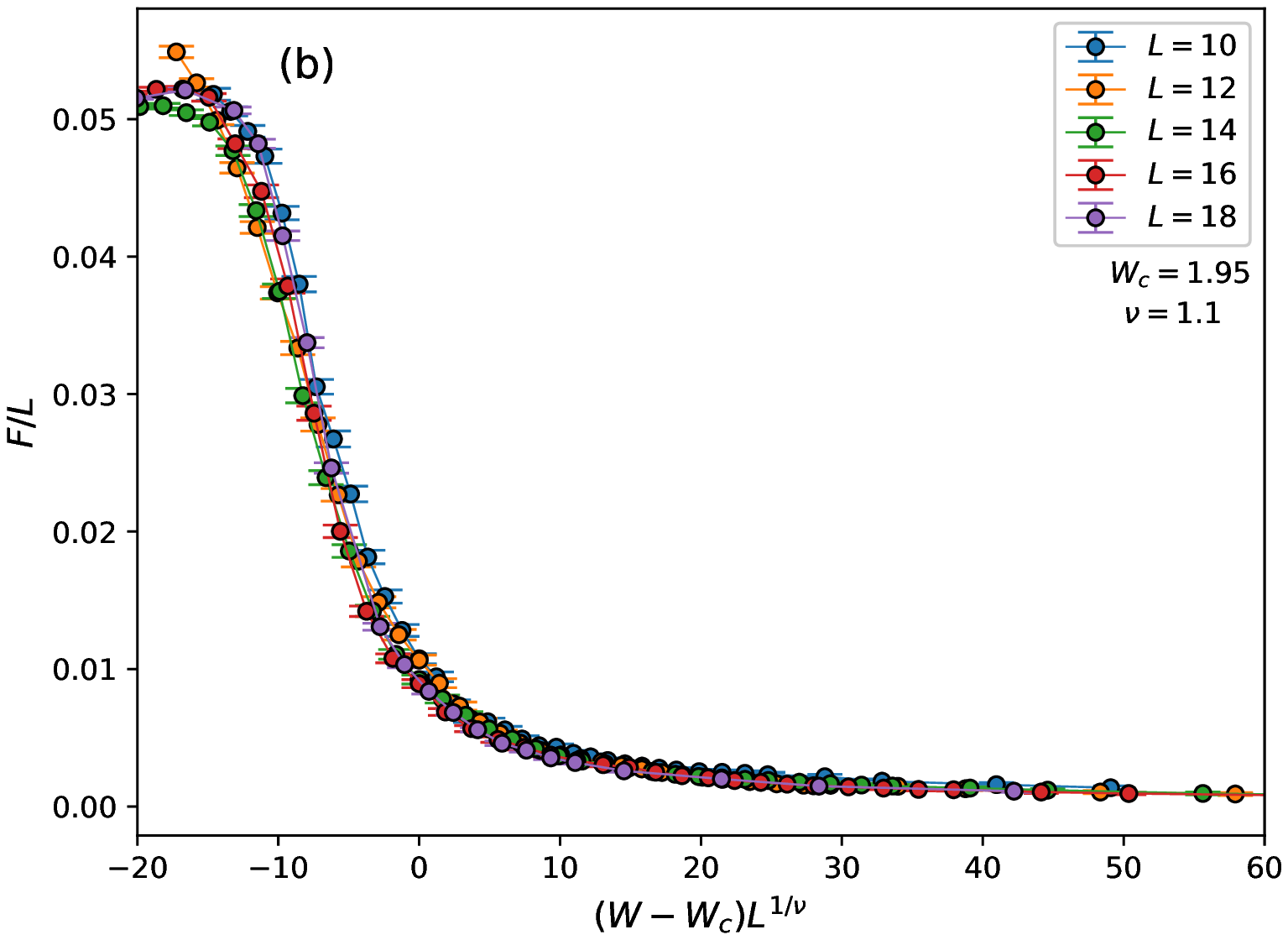}\\
	\caption{
		(Color online) (a) Finite-size scaling collapse for (a) entanglement entropy $S/S_T$ with $S_T$ as the Page value\cite{page1993}, and (b) the fluctuations of the half-system magnetization data in the quasiperiodic model for the system sizes $L=10-18$.  The critical disorder strength $W_{cl}\sim 1.85-1.95$ and scaling exponent $\nu \sim1.1$ are used to best collapse the data.}
	\label{fig2}
\end{figure}

We perform ED calculations to obtain energy eigenstates around the energy $E$ at a target energy density $\varepsilon$\cite{luitz2015} for systems with different number of sites $L=10-18$ in the total $S_z=0$ sector.  Specifically, for each quasiperiodic field configuration, we first calculate the ground state energy $E_0$ and the maximum energy $E_\text{max}$, which are used to define the target energy density $\varepsilon = (E-E_0)/(E_\text{max} -E_0)$.  We first locate the critical point for the MBL phase transition based on the entanglement entropy and the fluctuations of the half-system magnetization\cite{luitz2015}.  The Von Neumann entanglement entropy of a system partitioned in the middle, with reduced density matrix $\rho_A$, is given by $S=- \Tr (\rho_A \ln \rho_A)$.  We average the bipartite entanglement entropy over 30 ($L=10$) to 200 ($L=18$) eigenstates near target energy $E$ characterized by energy density $\varepsilon=0.5$, and over $1000$ quasifield  configurations by choosing random $\phi$ between $(0, 2\pi)$.  As shown in Fig. 1(a), we plot the ratio of entanglement entropy over the number of system sites $S/L$ for different systems at energy density $\varepsilon=0.5$ from $L=10$ to $18$ as a function of quasiperiodic field strength $W$.  As $W\to0$ we see $S/L$ increases with $L$ which approaches the Page value ($S/L \sim 0.5\ln(2)$ for large $L$ limit)\cite{page1993} following the volume law of the ergodic phase.  For larger $W$, $S/L$ approaches zero indicating area law entanglement and non-ergodic behavior where the MBL state is realized.  With varying $W$, all data points approximately cross each other around a critical value $W_{cl} \sim 1.85$.  We compare the entanglement entropy behavior with the bipartite fluctuations $F$ of the subsystem magnetization $S^z_A$~\cite{luitz2015,song2012}, which is defined as $F=\braket{{S^{z}_A}^2} - \braket{S^z_A}^2$ as shown in Fig. 1(b).  We see that $F/L$ increases on the small $W$ side, while it becomes vanishingly small on the larger $W$ side.  The $F/L$ curves for different $L$ approximately cross each other around the  critical field strength $W_{cl}\sim1.85$, consistent with the behavior of the entanglement entropy.  In fact, we see that there is an approximately proportional relationship between $S$ and $F$  for all $W$ region.  We also note that the crossing points between larger $L$ curves move towards the larger $W$ side.  This feature was also observed in a different model for quasiperiodic systems as well as for random disorder systems\cite{vedika2016, vedika2017}, which indicates the $W_{cl}$ we observed is a lower bound for the critical point of the dynamic quantum phase transition.

\begin{figure}[b]
	\vspace{10 pt}
	\includegraphics[angle=-90,width=0.49\linewidth]{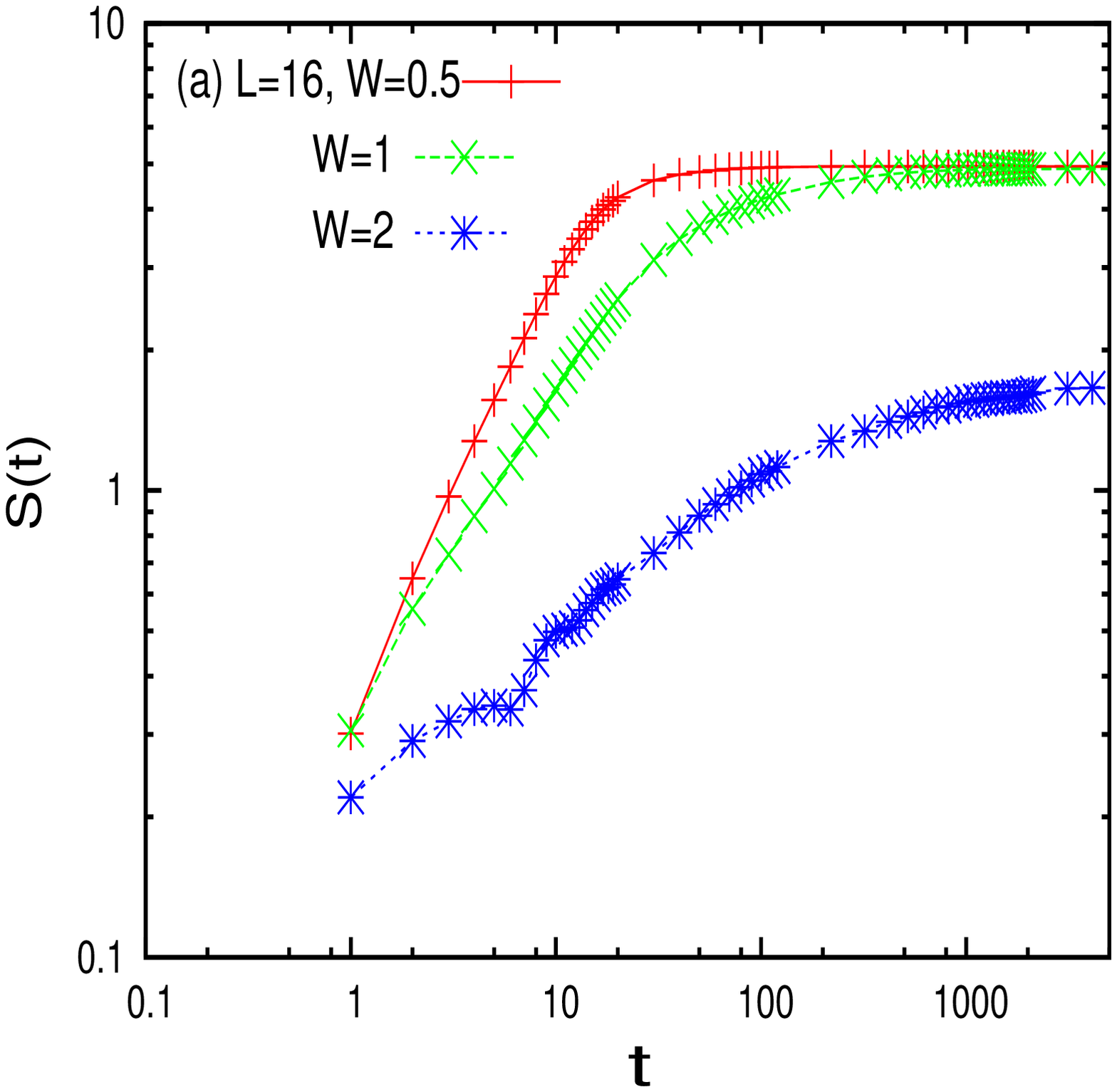}
	\includegraphics[angle=-90,width=0.49\linewidth]{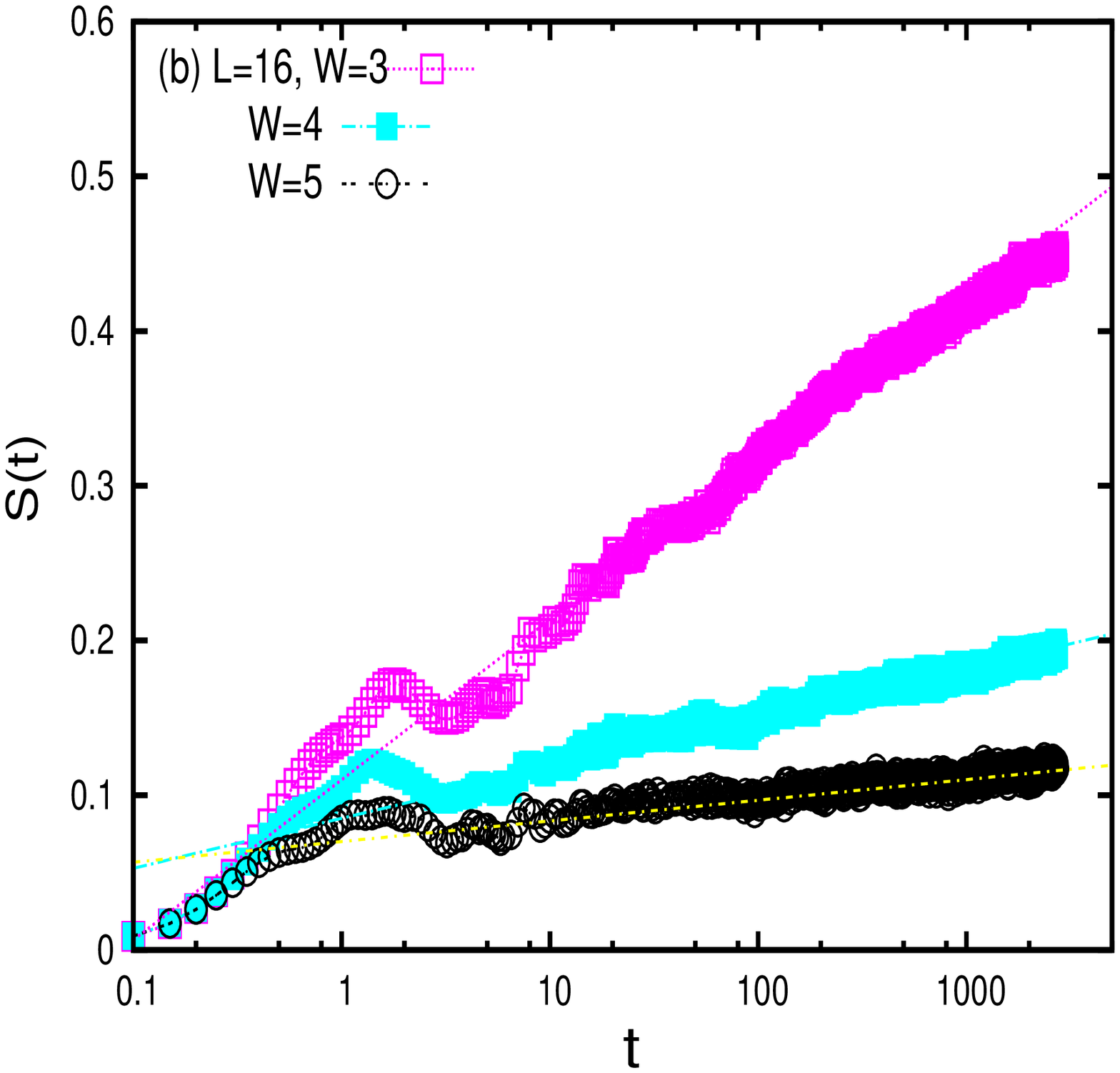}\\
	\caption{
		(Color online) (a) In this log-log plot, for systems with $W<W_c$ we observe power-law growth of entropy $S(t)$ which saturates at the $L=16$ Page value. (b) Semi-log plot of $S(t)$ with $W > W_c$ indicating logarithmic growth of $S(t)$. The error bars of data are at the same order as the size of symbols.}
	\label{fig3}
\end{figure}

\begin{figure*}[t]
	\centering
	\includegraphics[angle=-90,width=.30\linewidth]{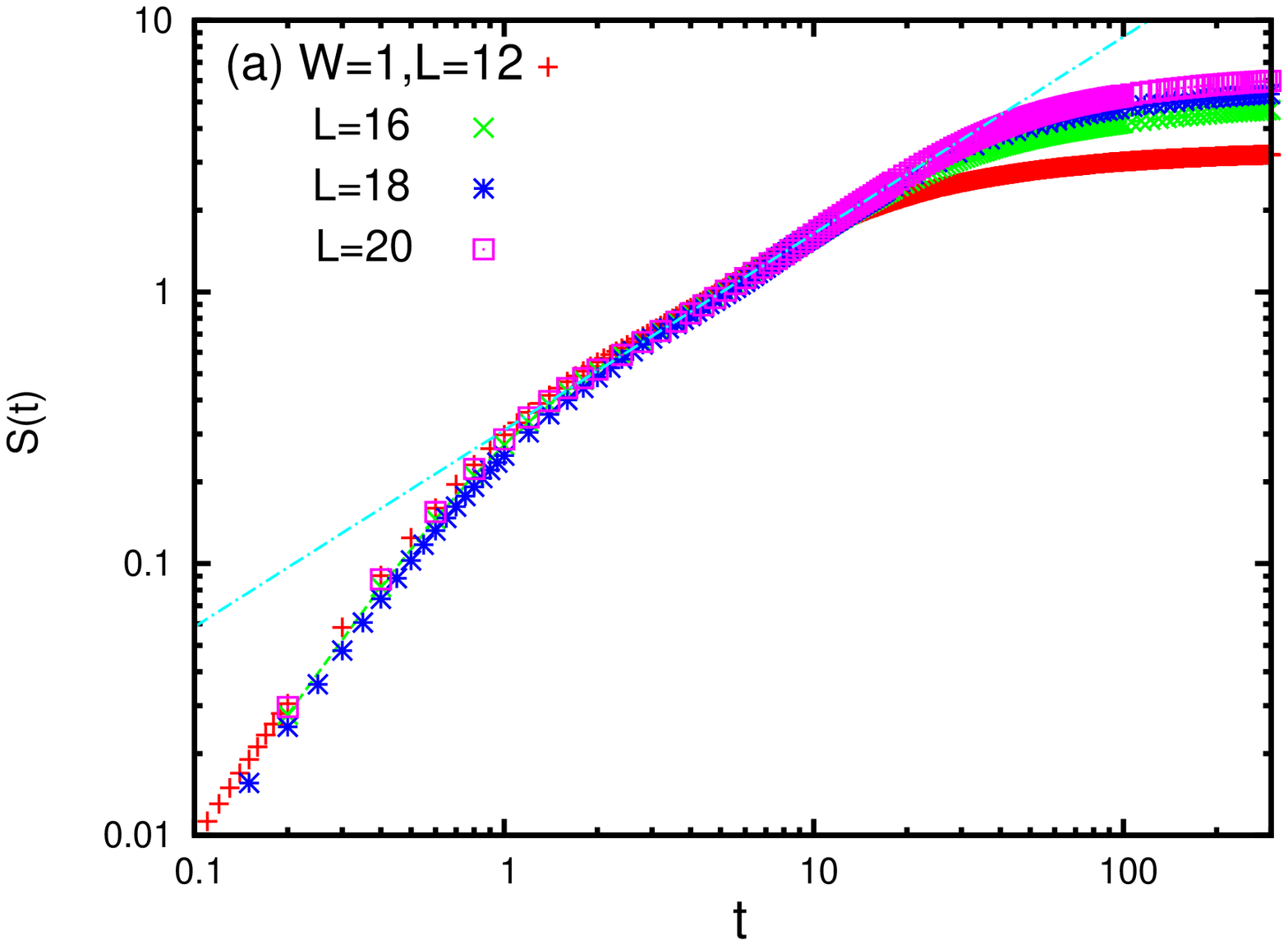}
	\includegraphics[angle=-90,width=.30\linewidth]{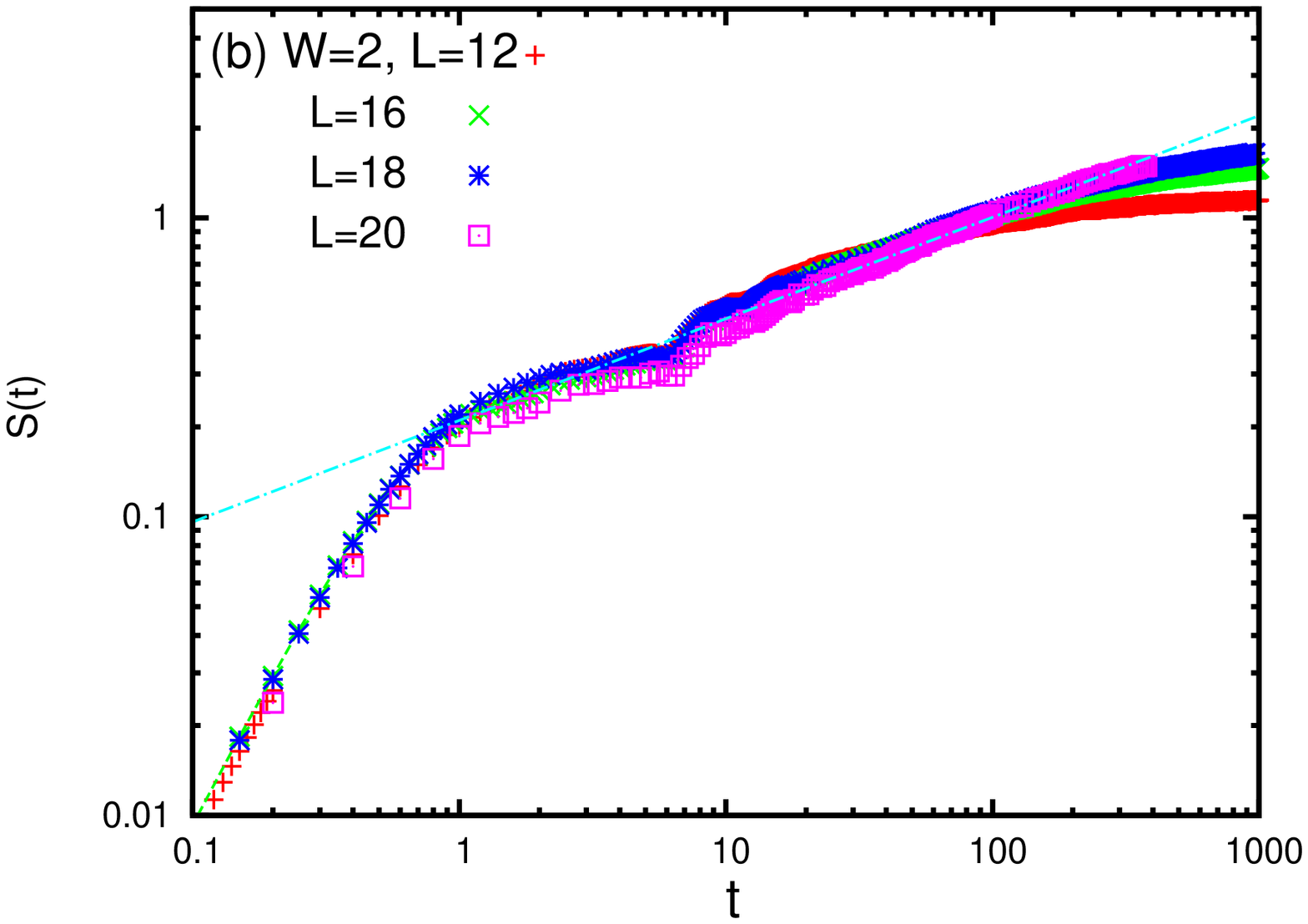}
	\includegraphics[angle=-90,width=.30\linewidth]{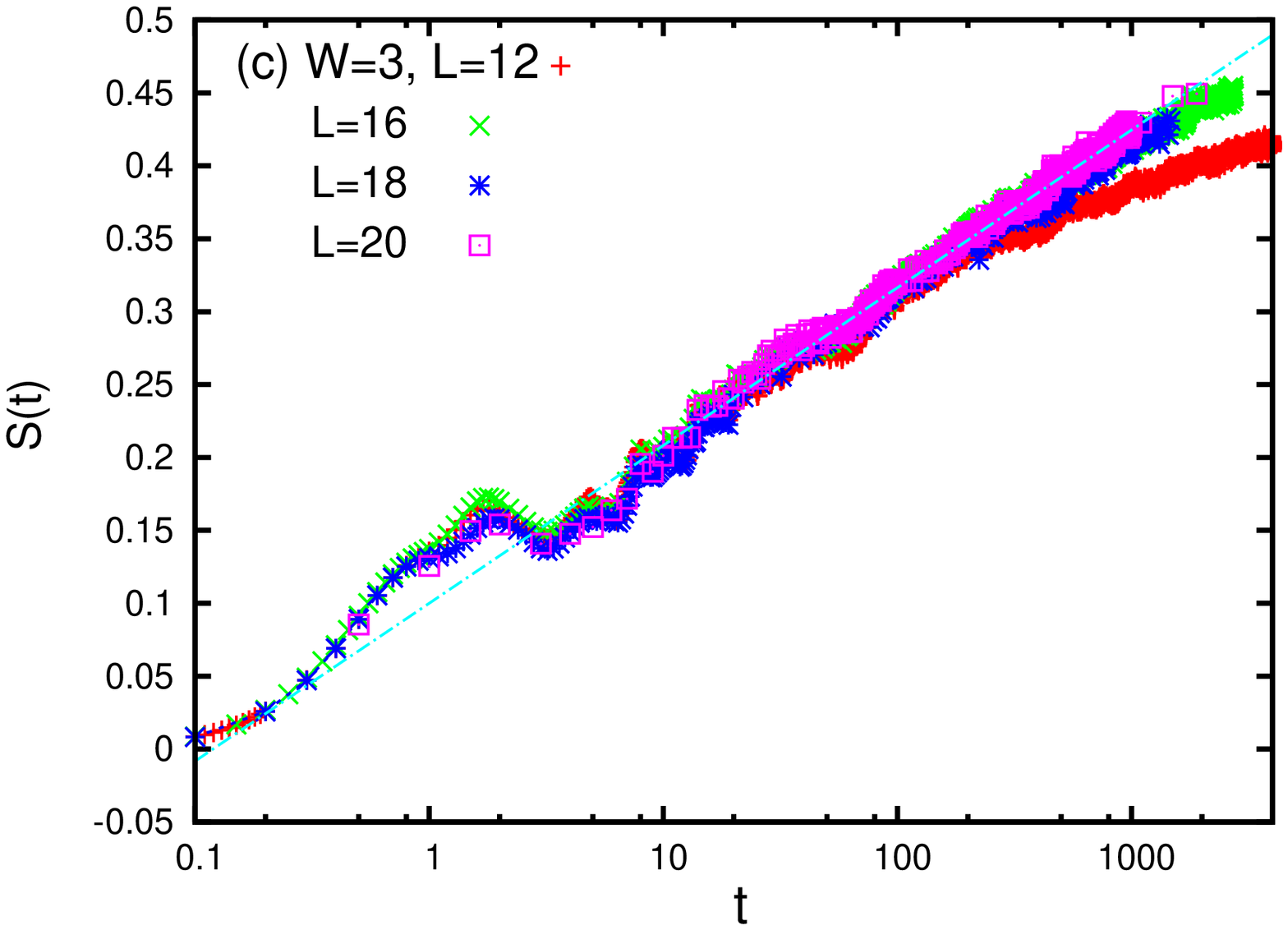}
	\caption{
		(Color online) (a) For system sizes ranging from $12$ to $20$, at $W=1$, we observe that $S(t)$increases rapidly until $t \sim 1$.  When $1<t<50$, $S(t)$ for all $L$ data fit into a straight line demonstrating robust power-law growth.  For larger $t$, we see that $S(t)$ saturates toward $\frac{L}{2} \ln{2}$,  consistent with the thermal entropy of the ergodic phase. (b) At $W=2$, for smaller system sizes, we observe that $S(t)$ grows slower than predicted by the power-law; on the other hand, $L=20$ results behave as expected for a thermal state and the growth of its entropy over time follows the power-law. (c) At $W=3$, we notice that all $S(t)$ plots fit to a straight line in the semi-log plot, indicating logarithmic growth for the MBL state.}
	\label{fig4}
\end{figure*}

We now analyze the finite-size scaling properties of the MBL transition for the quasiperiodic field model.  Crossing the quantum phase transition, we expect that the entanglement entropy ratio $S/S_T$ and the fluctuations of the half system magnetization over the system length $F/L$ should be a function of
$L/\xi\sim  L(W-W_c)^{\nu}$, where the correlation length $\xi$ has power-law divergence at the transition point with an exponent $\nu$.  Note that $S_T=0.5(L\ln(2)-1)$ is the saturated thermal value for the entanglement entropy of a finite size system\cite{page1993, vedika2017}.  As shown in Fig. 2(a-b), we find that these quantities for all system lengths can indeed be collapsed into one curve in a form $f((W-W_c)L^{1/\nu})$ by using the proper critical $W_c\sim 1.85-1.95$ and the scaling exponent $\nu\sim 1.1 \pm 0.1$, which give the best collapsing effect.  The obtained exponent $\nu$ is in agreement with the results of Khemani et al.\cite{vedika2017}, although the fitting for the $F/L$ shows much larger finite-size effect. 

\subsection{Time-evolution of quantum states}

We study the non-equilibrium quantum dynamics of the quasiperiodic systems after a global quantum quench.  Here we start by selecting a product state $\ket{\Psi(0)}=\ket{\sigma_1, \sigma_2, \ldots , \sigma_L}$ 
with an average energy close to the target energy determined by the energy density $\varepsilon=0.5$ at the time $t=0$ after the quench, where $\sigma_i=\pm$ represents the spin-z component $\pm 1/2$ (with $\hbar=1$) at site $i$.  The state at time $t$ can be obtained as $\ket{\Psi(t)} = e^{-iHt}\ket{\psi(0)}=e^{-iH\Delta t}\ket{\psi(t-\Delta t)}$.  We calculate the time evolution of an initial state $\ket{\Psi_0}$ based on a projection of the Hamiltonian to the Krylov space spanned by $\ket{\Psi_0}$, $H\ket{\Psi_0}$, \ldots, $H^n\ket{\Psi_0}$.  We calculate all eigenstates in this space to obtain the time-evolution operator\cite{luitz2015}.  Using a reasonably small time step $\delta t \sim 0.2/J$ allows for highly accurate results for the quantum state  with a small $n=30-60$.  All time-evolution results are being averaged over more than 500 quasiperiodic field configurations. 

We first discuss the general behavior of the entanglement entropy as a function of time.  On the small $W$ side shown in Fig. 3(a), we find that the entropy $S(t)$ exhibits power-law growth in time $t$ before it reaches the saturated value $\frac L 2 \ln2$ at long time limit in agreement with the ETH.  On the larger $W$ side, we find a much slower growth, which can be fit with a logarithmic growth function as shown in Fig. 3(b) for $W=3-5$.  We now analyze the finite-size scaling behavior of $S(t)$ for $L=12-20$.  For small $W=1$ as shown in Fig. 4(a), we find that the initial growth ($t\sim 1$) of $S(t)$ is very rapid  and system size independent.  For the intermediate time regime, $S(t)$ experiences power-law growth as demonstrated by the linear behavior in the logarithmic plots until the finite-size effect sets in.  With the increase of $L$, we find a wider time interval for the power-law growth of $S(t)$.  Interestingly, we see very similar behavior and a smaller window for power-law growth of $S(t)$ for $W=2$.  The power-law growth indicated by the straight line in the Fig.4(b) is clearest for larger system size $L=20$.  This is a strong indication that the $W=2$ is in the thermal phase consistent with the moving of the crossing point toward larger $W$ with the increase of $L$ observed in Fig. 1.  We then look into $S(t)$ at $W=3$ as shown in Fig. 4(c) where we observe that for small $t$, $S(t)$ grows rapidly  while the initial product state evolves to a superposition state for $t\sim 1$, which is then followed by some oscillations of $S(t)$.  With further increase of $t$, we find a logarithmic growth of $S(t)$ for a time range of more than two orders of magnitude.  The range of $t$ for logarithmic growth of $S(t)$ becomes larger with the increase of $W$.  These results confirm an MBL phase with similar behavior to the random field case studied by Luitz et. al\cite{luitz2015}.  

Now we turn to spin correlations during the time evolution.  We start from the product state $\ket{\Psi(0)}$ where the $S_i^z$ on each site $i$ is $\pm 1/2$ while the total $S_z$ of all sites is zero.  We define the following time correlator for $\sigma_z$ as 
\begin{equation}
	I(t)=\frac{4}{L} \sum_{j=1}^L \braket{\Psi(0)|S_j^z(0)S_j^z(t)|\Psi(0)} \text{,} \label{eq:imb} 
\end{equation}
which detects the total imbalance of spin-z component.  As shown in Fig.5(a), we find a systematic change of the properties of $I(t)$ as $W$ is varied.  For smaller $W=0.5$ and $1$, we see that the long time behavior of imbalance $I(t)$ is dominated by power-law decay $t^{-\zeta}$, and at the large $t$ limit $S_i^z$ on a site becomes uncorrelated with the initial condition and $I(t)$ approaches zero.  For intermediate $W=1.5$ and $2$, a similar power-law behavior is obtained with a much smaller decay power $\zeta$, indicating the longer time scale required to approach equilibrium spin correlations for these thermal states near the transition point to the MBL phase.  On the MBL side with $W=3$ and $4$, we see that the $I(t)$ is near constant at large $t$ limit with a near vanishing decay exponent ($\zeta \sim 0$).
In Fig. 5(b), we show the decay exponent $\zeta$ as a function of $W$, where we find that the critical point for the transition to MBL phase is close to $W_c \sim 2.5\pm 0.04$ consistent with the conjecture that $W_{cl}\sim 1.85$ is only the lower bound of the critical point, though for given range of system sizes $L=10-18$ it does give the best collapsing of the finite sizes entropy and fluctuation data (for highly excited eigenstates) as shown in Fig. 2.

\begin{figure}[b]
	\centering
	\begin{tikzpicture}
	\node (o) at (0, 0) {};
	\node (a) [inner sep=0pt, left] at (o.west) {\includegraphics[width=0.45\linewidth]{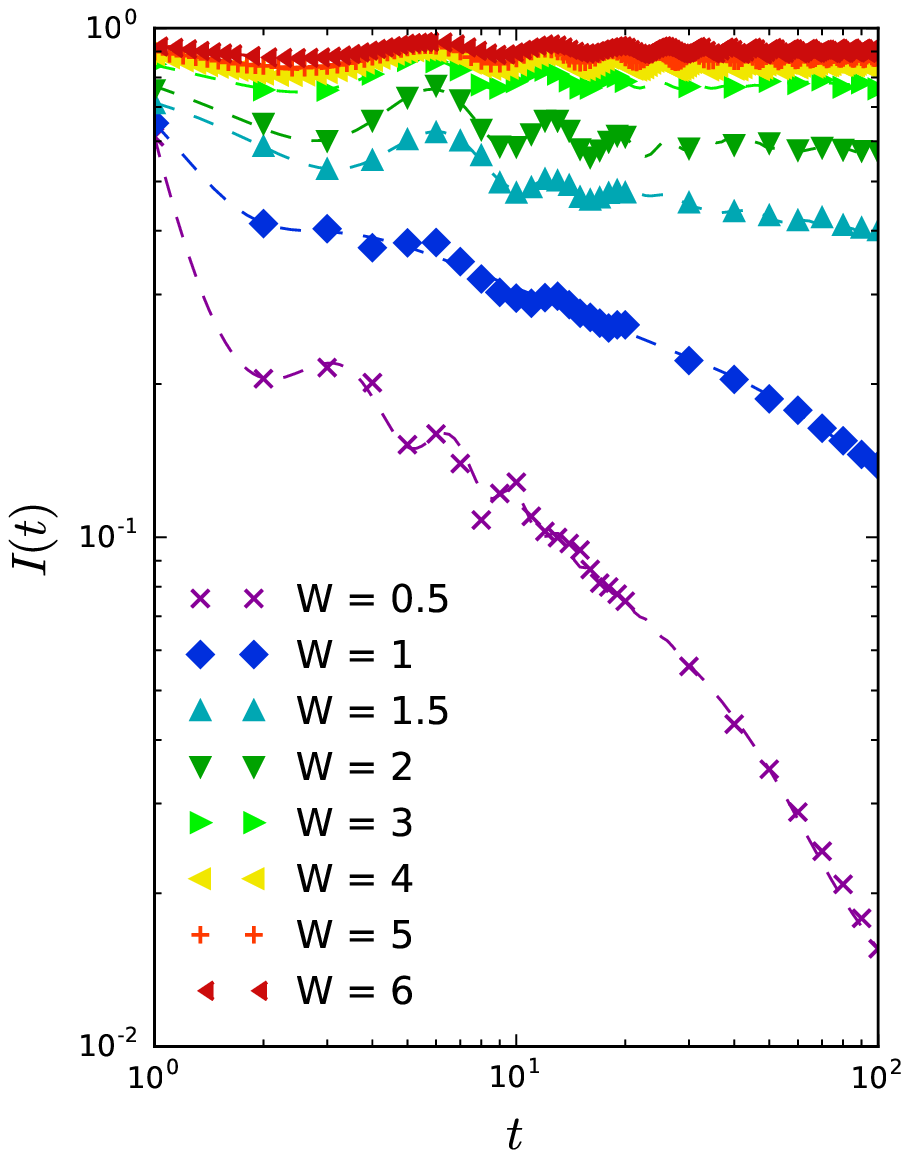}};
	\node (b) [inner sep=0pt, right] at (o.east) {\includegraphics[width=0.45\linewidth]{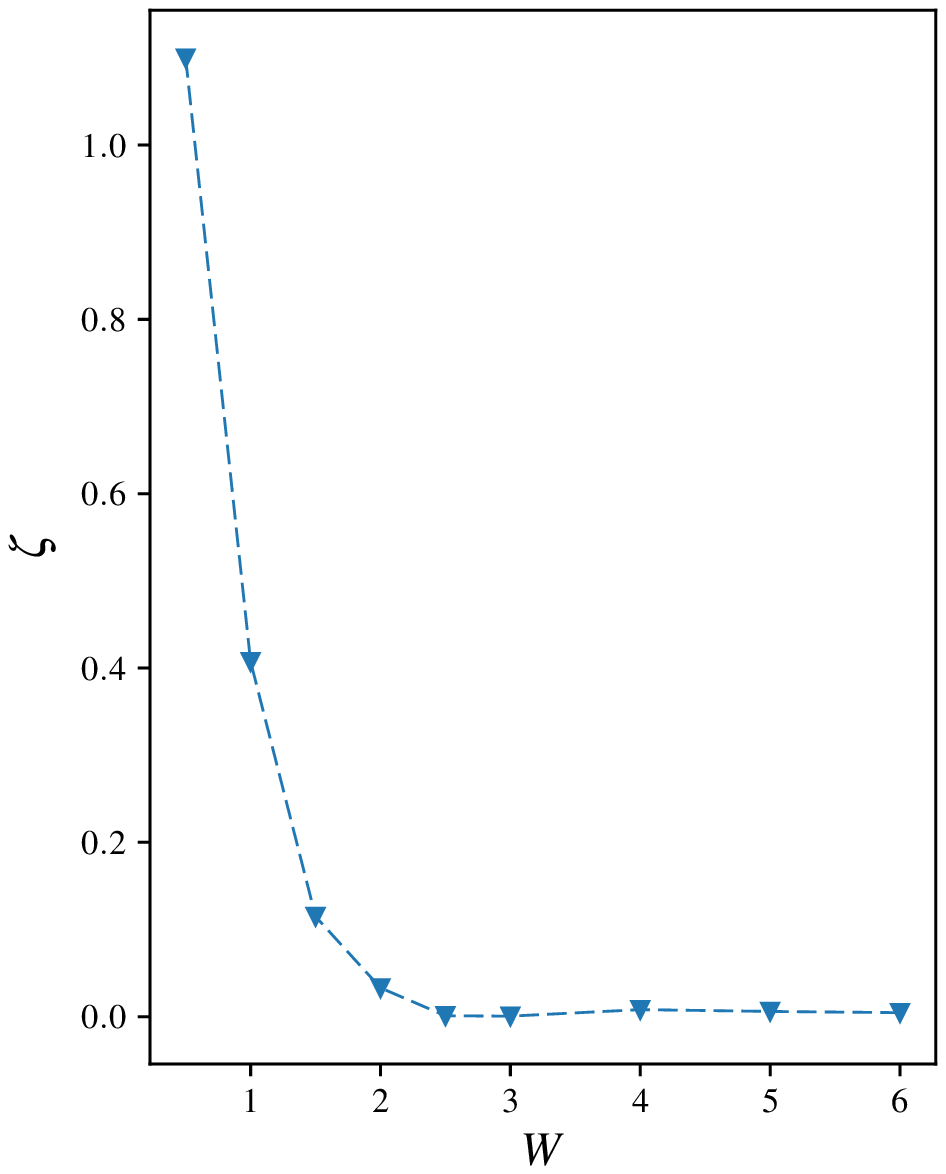}};
	\node [right] at (a.north west) {(a)};
	\node [right] at (b.north west) {(b)};
	\end{tikzpicture}
	\includegraphics[angle=-90,width=.95\linewidth]{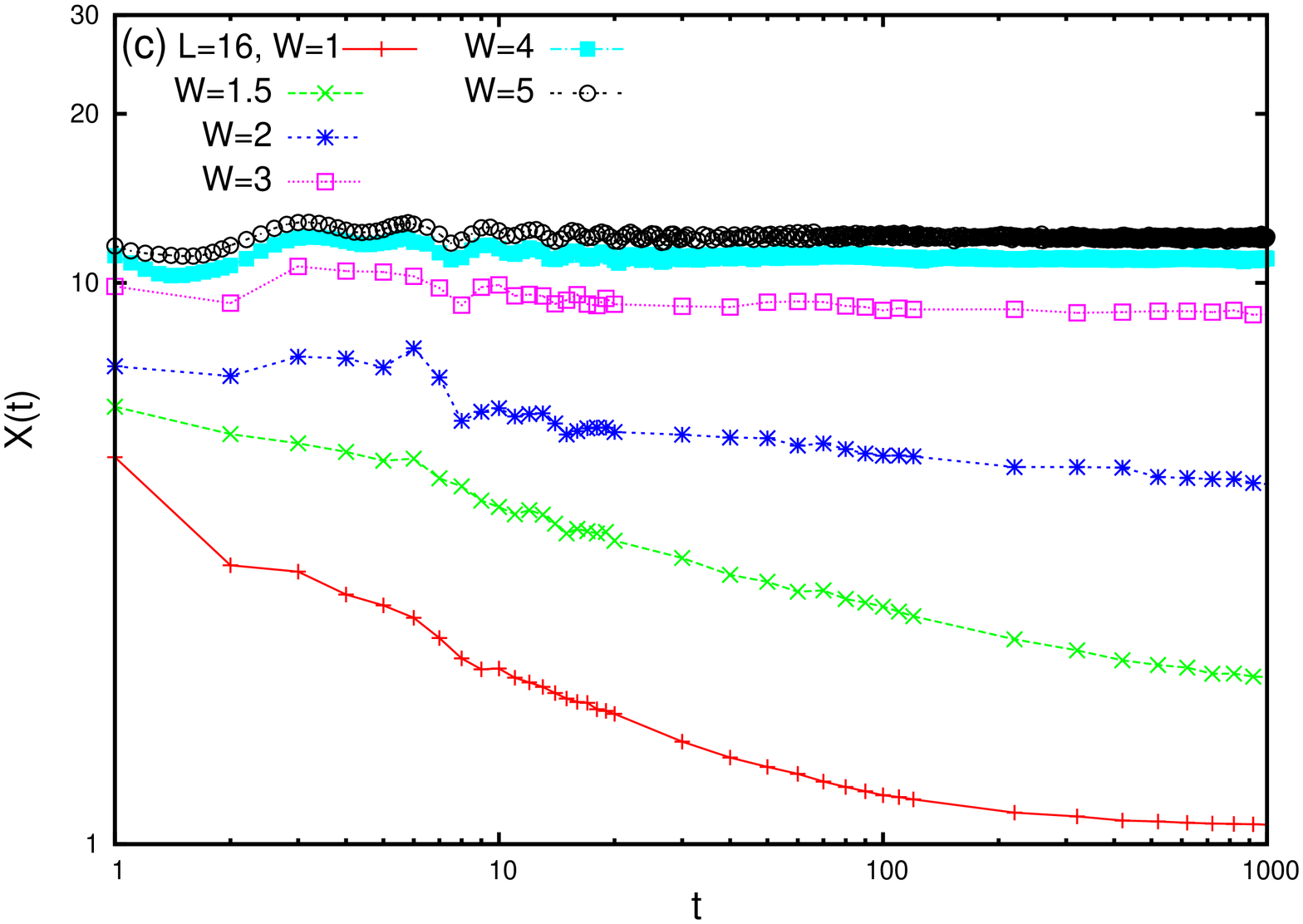}\\
	\caption{
		(Color online) (a) Imbalance $I(t)$ for a $L=16$ system with $W=0.5-6$.  At small values of $W$, we notice the $I(t)$ of the system decays rapidly; however, when $W$ is increased beyond $W>2$, $I(t)$  ceases decaying and remains at a certain level.  This is consistent with the MBL behavior where the initial values of the local observables for each site $i$ is preserved. (b) Fitting parameter $\zeta$ for the power-law decay exponent as a function of $W$. (c) Spin glass order as defined in eq.\ref{eq:sgo}. At small values of $W$, correlations between spins are short ranged and $\chi(t)$ decays to $1$ over time; at large values of $W$, $\chi(t)$ remains close to its initial value even at very long time, further corroborating our assertion of a quantum phase transition.}
	\label{fig5}
\end{figure}

For comparison, we also study spin glass order\cite{kjall2014} for the MBL phase.  The spin-flip from the Heisenberg term will create domain walls.  If the domain walls are confined together, a spin-glass order can develop.  We define the spin glass order parameter 
\begin{equation}
	\chi=\frac 1 L \sum_{i,j=1}^L \braket{\Psi(t)|S_i^zS_j^z|\Psi(t)}^2 \text{,} \label{eq:sgo}
\end{equation}
which can diverge with $L$ in the spin-glass ordered phase.  As shown in Fig.5(c), we see behavior very similar  to $I(t)$.  For smaller $W=1-2$, we see that $\chi(t)$ decreases with $t$ in power-law fashion, while it maintains a large value in the long time limit for larger $W>2$.  Our results indicate a jump of the spin glass order at the thermal to MBL transition.  In Fig.6, we see that both $I(t)$ and $\chi(t)/L$ show very weak size dependence at $W=3$ and remain nonzero at long time and large system size limits, which fully establish the robustness of the MBL phase.  

\begin{figure}[h]
	\centering
	\includegraphics[angle=-90,width=0.49\linewidth]{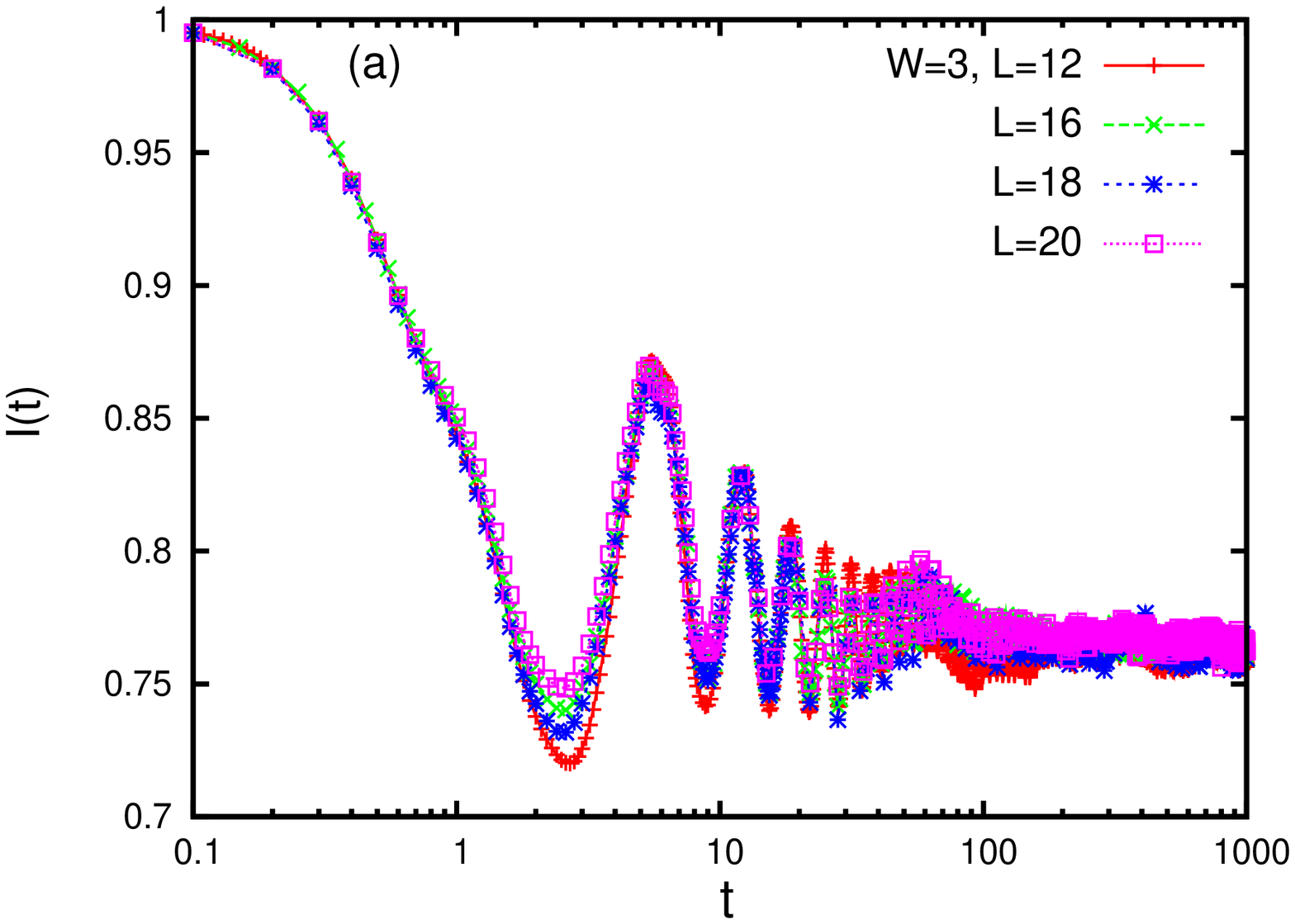} 
	\includegraphics[angle=-90,width=0.49\linewidth]{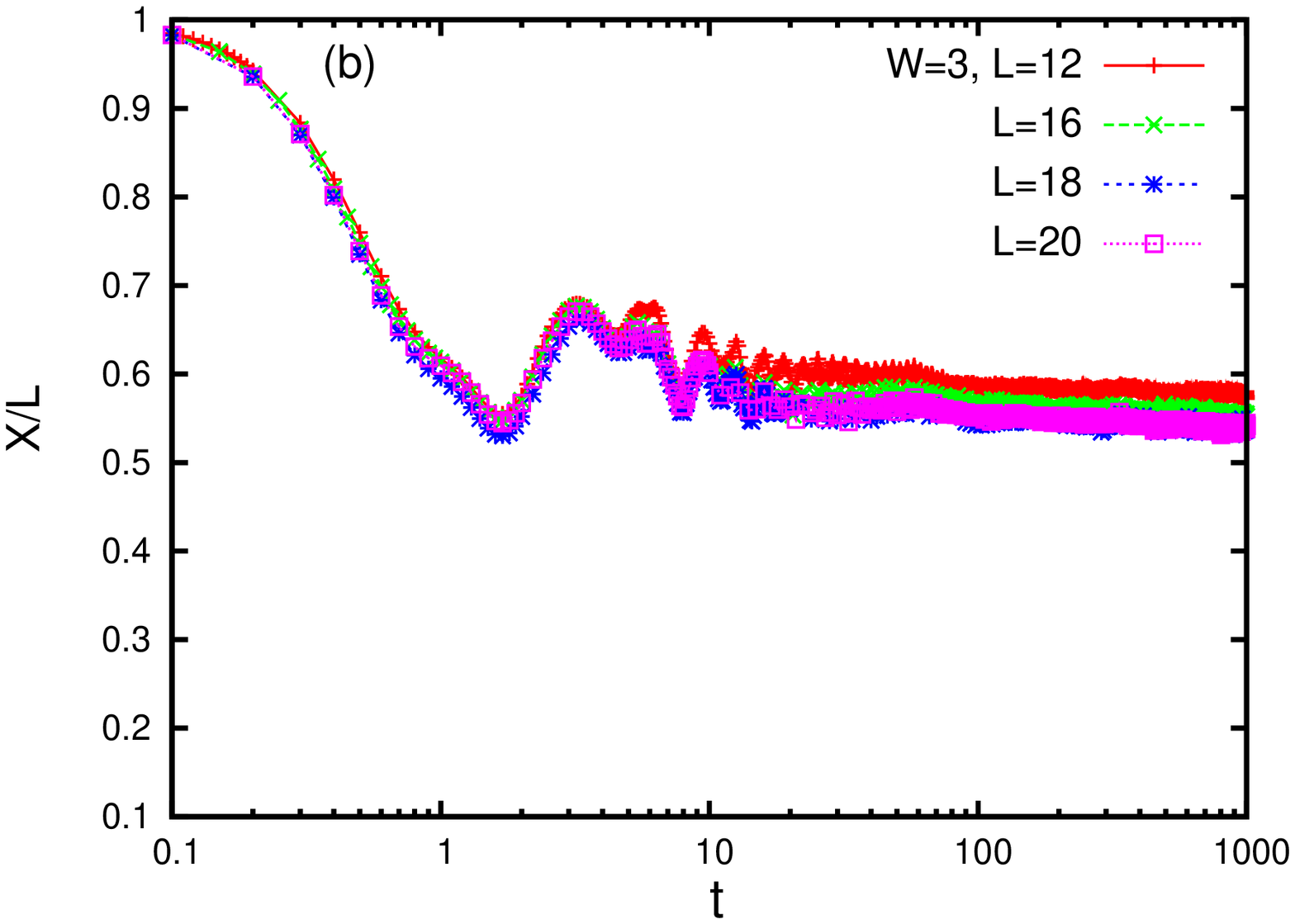}\\
	\caption{
		(Color online) (a) Imbalance $I(t)$ for a $L=12-20$ systems at $W=3$.  $I(t)$ is insensitive to system size $L$.  It shows initial oscillation at shorter times, and it stays at nonzero value at the long time limit.  (b) Spin glass order $\chi(t)/L$ saturates to finite nonzero value indicating the divergent behavior with $L$.}
	\label{fig6}
\end{figure}

\section{Summary and Discussions}

We have studied many-body localization and quantum phase transitions in spin chain systems in the presence of quasiperiodic fields.  Based on the entanglement entropy and the fluctuation of the half system magnetization studies, we find the lower bound of the critical field strength $W_{cl}$ for the dynamic quantum phase  transition from the thermal phase to the MBL phase  driven by the quasiperiodic fields to be on the order of $1.85$. Interestingly, for $W$ just above $W_{cl}$, we find that the entanglement entropy following a global quench grows with time under the power law, consistent with the behavior of the thermal phase.  From the scaling behavior of the spin imbalance and spin glass order, we identify the divergent spin glass order for $W\geq 3$.  Overall, the finite size effect in such a quasiperiodic system turns out to be not too important as the scaling behavior in the intermediate regime near the transition appears to be showing either thermal behavior ($W \sim 2 $) or MBL characteristics ($W\sim 3$), suggesting the best estimate of the transition point to be $W_c=2.5\pm 0.4$.  Our results provide quantitative understanding of the effect of quasiperiodic fields, which are more efficient in driving MBL physics than random fields due to importance of the rare Griffiths regions in random field models.  It would be very interesting to further identify the phase diagram for quantum states at different energy densities and reexamine the possibility of the existence of the mobility edge\cite{newpaper} in such systems based on large scale
density matrix renormalization studies.  Another interesting direction is to explore the MBL phase transition in ladder systems with coupled spin chains\cite{baygan2015}, which provides information about the MBL physics in two dimensions.

{\bf Acknowledgments} --- D.N.S. thanks  Vedika Khemani and David A. Huse for collaborations on a related work.  This work is supported by US National Science Foundation  Grants PREM DMR-1205734 (M.L., T.R.L.) and DMR-1408560 (D.N.S., S.P.L.).

M.L. and T.R.L. contributed equally to the preparation of
this work.


\end{document}